\documentclass[10pt, twocolumn, pre, aps, superscriptaddress, showpacs]{revtex4-1}
\usepackage{amsmath, graphicx, subfigure, tikz}
\begin{document}
    \title{Phase Transitions of the Variety of Random-Field Potts Models}
    \author{Alpar T\"urko\u{g}lu}
    \affiliation{Department of Electrical and Electronics Engineering, Bo\u{g}azi\c{c}i University, Bebek, Istanbul 34342, Turkey}
    \author{A. Nihat Berker}
    \affiliation{Faculty of Engineering and Natural Sciences, Kadir Has University, Cibali, Istanbul 34083, Turkey}
    \affiliation{Department of Physics, Massachusetts Institute of Technology, Cambridge, Massachusetts 02139, USA}

    \begin{abstract}
The phase transitions of random-field $q$-state Potts models in $d=3$ dimensions are studied by renormalization-group theory by exact solution of a hierarchical lattice and, equivalently, approximate Migdal-Kadanoff solutions of a cubic lattice.  The recursion, under rescaling, of coupled random-field and random-bond (induced under rescaling by random fields) coupled probability distributions is followed to obtain phase diagrams.  Unlike the Ising model $(q=2)$, several types of random fields can be defined for $q \geq 3$ Potts models, including random-axis favored, random-axis disfavored, random-axis randomly favored or disfavored cases, all of which are studied.  Quantitatively very similar phase diagrams are obtained, for a given $q$ for the three types of field randomness, with the low-temperature ordered phase persisting, increasingly as temperature is lowered, up to random-field threshold in $d=3$, which is calculated for all temperatures below the zero-field critical temperature.  Phase diagrams thus obtained are compared as a function of $q$. The ordered phase in the low-$q$ models reaches higher temperatures, while in the high-$q$ models it reaches higher random fields.  This renormalization-group calculation result is physically explained.
    \end{abstract}
    \maketitle

\section{Introduction: The Variety of Random-Field Potts Models}

Quenched randomness strongly and distinctively affects phases and phase boundaries.  Random bonds change the critical exponents of second-order phase transitions if the nonrandom specific heat critical exponent $\alpha$ is positive \cite{Harris,BerkerCrit}.  In fact, the nonrandom specific heat critical exponent $\alpha$ is proportional to the crossover exponent from nonrandom criticality \cite{AndelmanBerker}.  On the other hand, $\alpha$ is not proportional to the crossover exponent from random criticality \cite{AndelmanBerker,BerkerCrit}. Random bonds convert first-order phase transitions to second-order phase transitions, with even infinitesimal randomness in $d=2$ dimensions \cite{AizenmanWehr,AWerr,HuiBerker,erratum,BerkerConv} and after a threshold amount of randomness in $d \geq 2$ \cite{HuiBerker,erratum,BerkerConv}. The random introduction of competing bonds induces a new phase, the spin-glass phase, with the characterisic signature of chaos under scale change \cite{McKayChaos,McKayChaos2,BerkerMcKay}. Random fields eliminate ordered phases in low dimensions \cite{ImryMa}.  In fact, for the Ising model, an extended experimental controversy on the lower-critical dimension (at and below which no ordering occurs) being $d_c = 2$ \cite{Jaccarino,Wong} or $d_c = 3$ \cite{Birgeneau} was eventually settled for $d_c = 2$, as supported by renormalization-group calculations \cite{BerkerRF,Machta,Falicov}.  Recent studies of the random-field Ising model are in Refs. \cite{Fytas1,Fytas2,Fytas3}

The previously controversial Ising model is the $q=2$ state case of the Potts models, defined by the Hamiltonian
\begin{equation}
- \beta {\cal H} =
\sum_{\left<ij\right>} J\delta(s_i,s_j) \,,
\end{equation}
where $\beta=1/k_{B}T$, at site $i$ the spin $s_{i}=1,2,...,q$ can
be in $q$ different states, the delta function $\delta(s_i,
s_j)=1(0)$ for $s_i=s_j (s_i\neq s_j)$, and $\langle ij \rangle$
denotes summation over all nearest-neighbor pairs of sites.  We now include random fields:
\begin{equation}
- \beta {\cal H} =
\sum_{\left<ij\right>} [J\delta(s_i,s_j) + H_i\delta(s_i,u_i)+H_j\delta(s_j,u_j)]\,,
\end{equation}
where $u_{i}=1,2,...,q$ is frozen randomly at each site $i$ with local field $H_i$.  For the Ising case of $q=2$, a random field $H_i$ on the spin $s_i$ simultaneously favors one of the two states and disfavors the other state, independently of the sign of $H_i$.  For the $q \geq 3$ Potts models, on the other hand, we can distinguish three different random-field models: (1) For $H_i = H > 0$, the random field at each site $i$ favors one random state $u_i$ and disfavors the other $q-1>1$ states; (2) for $H_i = - H < 0$, the random field at each site $i$ disfavors one random state $u_i$ and favors the other $q-1>1$ states; (3) for $H_i = \pm H$ randomly, the random field at each site $i$ randomly favors or disfavors one random state $u_i$ and, respectively, disfavors or favors the other $q-1>1$ states.  These different random-field models correspond to random-axis favored, random-axis unfavored, random-axis randomly favored or disfavored models.  We have solved all three of these random-field models in $d=3$ dimensions.  Under renormalization-group transformation, the system maps onto a renormalized system in which at a given site a random distribution of fields towards all Potts states exists and is included in the renormalized quenched probability distribution.  Previous studies of the random-field Potts models are in Refs. \cite{Nishimori,Aharony,Reed,Binder1,Binder2,Binder3,Weigel}

\section{Method: Global Renormalization-Group Theory of Quenched Probability Distributions}
Detailed phase diagrams for systems with quenched randomness are obtained by global renormalization-group theory, where the quenched coupled probability distribution of the interactions renormalizes under scale change and is followed as trajectories of the quenched probability distribution function \cite{AndelmanBerker,Falicov}.  We solve our systems as an exact solution of a $d=3$ hierarchical model \cite{BerkerOstlund,Kaufman1,Kaufman2}, which is equivalently the Migdal-Kadanoff approximation \cite{Migdal,Kadanoff} for the cubic lattice (Fig. 1).  As seen in Fig. 1(b), hierarchical models are models obtained by self-imbedding a graph \textit{ad infinitum} \cite{BerkerOstlund,Kaufman1,Kaufman2}.  The exact renormalization-group solution of the hierarchical model is obtained in the reverse direction, by summing over the internal spins of the innermost graphs at each renormalization-group step.  The Migdal-Kadanof approximation, on the other hand, is an intuitive approximation in which, at each renormalization-group step, bonds are moved and decimation is done, as in Fig. 1(a).  The renormalization-group recursion relations for the hierarchical model and the Migdal-Kadanoff approximation in Fig. 1 are identical, which ascertains that the Migdal-Kadanoff approximation is a physically realizable, robust approximation.  Thus, this hierarchical-model/Migdal-Kadanoff approach correctly yields the lower-critical dimensions of Ising \cite{JoseKadanoff}, Potts \cite{BerkerOstlund}, XY models \cite{JoseKadanoff,BerkerNelson}, the low-temperature critical phases of the antiferromagnetic Potts \cite{BerkerKadanoff1,BerkerKadanoff2,Saleur} and $d=2$ XY models \cite{JoseKadanoff,BerkerNelson}, the lower-critical dimensions of the spin-glass \cite{Demirtas,Atalay} and random-field Ising \cite{Falicov} models, chaos under rescaling in spin glasses \cite{Ilker2}, the experimental phase diagrams of surface systems \cite{BOP}, the phase diagrams of high-temperature superconductors \cite{Hinczewski}, etc. Other physically realizable approximations have also been used in studies of polymers \cite{Flory,Kaufman}, disordered alloys \cite{Lloyd}, and turbulence \cite{Kraichnan}.  Recent works using exactly soluble hierarchical models are in Refs.
\cite{Myshlyavtsev,Derevyagin,Shrock,Monthus,Sariyer,Ruiz,Rocha-Neto,Ma,Boettcher5}.
\begin{figure}[ht!]
\centering
\includegraphics[scale=0.4]{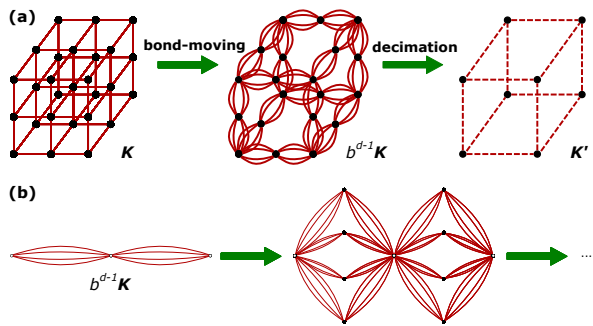}
\caption{(a) Migdal-Kadanoff approximate renormalization-group
transformation for the $d=3$ cubic lattice with the length-rescaling
factor of $b=2$. (b) Construction of the $d=3, b=2$ hierarchical
lattice for which the Migdal-Kadanoff recursion relations are exact.
The renormalization-group solution of a hierarchical lattice
proceeds in the opposite direction of its construction. From \cite{BerkerOstlund,ArtunBerker}.}
\end{figure}

The renormalization of the quenched coupled probability distribution is done by \cite{AndelmanBerker,Falicov}
\begin{equation}
P'(\textbf{K}'_{i'j'}) =
\int{\left\{\prod_{ij}^{i'j'}d\textbf{K}_{ij}P(\textbf{K}_{ij})\right\}\delta(\textbf{K}'_{i'j'}
- \textbf{R}(\left\{\textbf{K}_{ij}\right\}))}, \label{eq:conv}
\end{equation}
where $\textbf{K}_{ij} \equiv \{K_{ij}(s_i,s_j)\}$ is the general quenched-random nearest-neighbor interaction matrix occurring in the generalization of Eq.(2) induced by the renormalization-group transformation,
\begin{equation}
- \beta {\cal H} =
\sum_{\left<ij\right>} K_{ij}(s_i,s_j)\,,
\end{equation}
where, with no loss of generality, for each $(i,j)$, the same constant has been subtracted from each element in $K_{ij}$, so that the maximal element is 0 and all other elements are negative. Primes refer to the renormalized system.  $\textbf{R}(\{\textbf{K}_{ij}\})$ represents the local recursion relation, obtained by summing over the internal spins of the graph of the hierarchical model or, equivalently, by bond-moving and decimation in the Migdal-Kadanoff approximation.  The different asymptotic renormalization-group flows of $P(\textbf{K}_{ij})$ identify the different phases and phase transitions of the system:  All trajectories originating within the ordered phase flow to the stable fixed point (phase sink) of the ordered phase with $K_{ij}=0,-\infty$ for $i=j,\,i\neq j$, respectively.  All trajectories originating within the disordered phase flow to the stable fixed point (phase sink) of the disordered phase with $K_{ij}=0$ for all $i,j$.  The trajectories originating on the boundary between the two phases flow to an unstable strong coupling fixed point with $K_{ij}=0,-\infty$ for $i=j,\,i\neq j$, respectively.

Equation (3) is effected as follows: For the calculations for given $q$, a distribution of $qN$ nearest-neighbor $<ij>$ transfer matrices
\begin{equation}
E_{ij}(s_i,s_j) = e^{K_{ij}(s_i,s_j)}\,,
\end{equation}
is constructed at the beginning of the trajectory, from Eq.(2) and the type of the Random-Field Potts Model (1), (2), or (3).  At the beginning of the trajectory, the fields $(H_i,u_i,H_j,u_j)$ entering $K_{ij}$ are chosen randomly according to the rules of the random-field model.  This distribution is symmetrized with respect to $<ij>$ by augmenting it with the transpose matrices and with respect to $q$ by augmenting it with the matrices where simultaneously each row and column are cyclically augmented.  These symmetrization operations are done to cure any small asymmetry artificially introduced by the random numbers.  Thus, at the end of the symmetrization operations, we have a quenched random distribution of $2q^2N$ transfer matrices. The factor $q^2$ in the number of transfer matrices $2q^2N$ is included to fully reflect the multiplicity $q^2$ of random-field directions $(u_i,u_j)$ in Eq.(2).

\begin{figure}[ht!]
\centering
\includegraphics[scale=0.35]{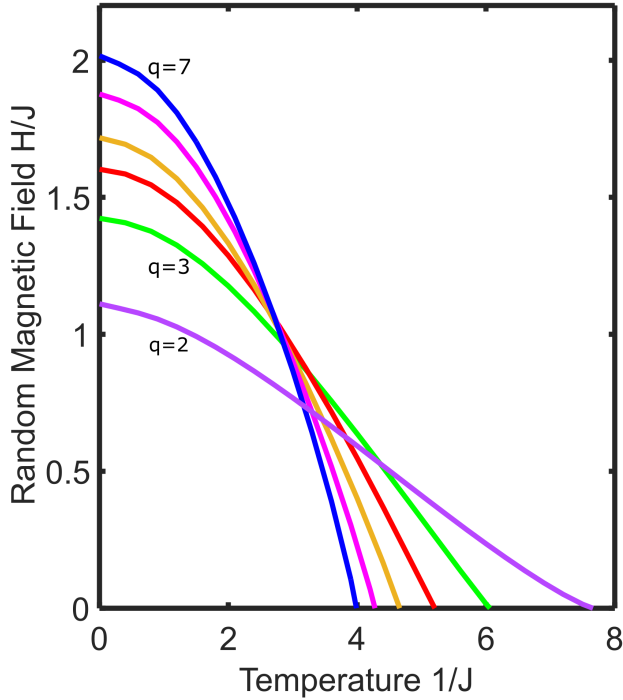}
\caption{Calculated phase diagrams of the random-axis randomly favored or disfavored random-field Potts models in $d=3$, for the number of states $q = 2,3,4,5,6,7$.  The curves are for decreasing $q$ along the increasing temperature axis and along the decreasing random-field strength axis. The $q=2$, namely Ising case, is as in Ref.\cite{Falicov}.  Thus, interestingly, the ordered phase in the low-$q$ models reaches higher temperatures, while in the high-$q$ models it reaches higher random fields.  A physical explanation is given in the text.}
\end{figure}
\begin{figure}[ht!]
\centering
\includegraphics[scale=0.35]{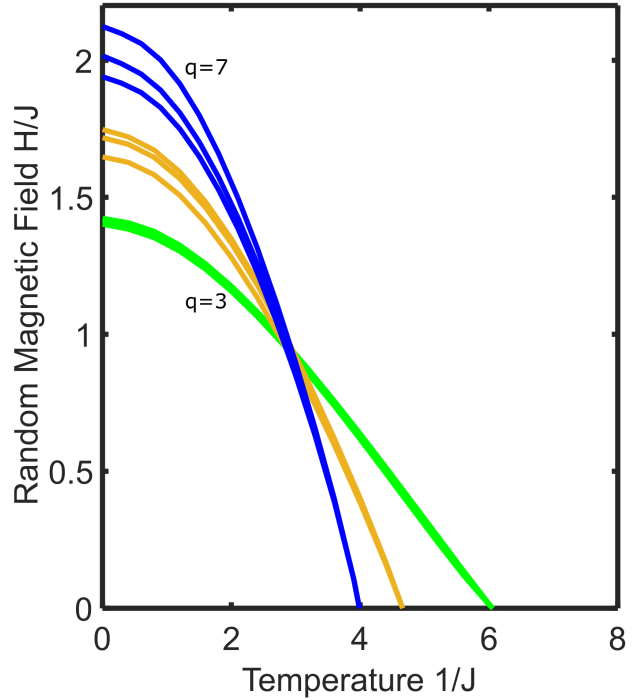}
\caption{Calculated phase diagrams for the differently defined random-field Potts models: (1) random-axis favored, (2) random-axis unfavored, (3) random-axis randomly favored or disfavored models.  The three differently defined models are not expected to have superimposed phase diagrams for a given $q$. This is seen in this figure for $q = 3,5,7$, where Models (1), (2), (3) are not superimposed, but very closely track each other.  The triplet of curves (for the three differently defined models within each triplet) are for decreasing $q$ along the increasing temperature axis and along the decreasing random-field strength axis.  The triplets are, for increasing field at low temperatures, for Models (1,2,3), (1,3,2),(1,3,2) for $q = 3,5,7$ respectively.  The three curves in the $q=3$ triplet are not distinguishable on the scale of this figure.  We have also calculated the three Model phase diagrams for $q = 4,6$, not shown in this figure for congestion reasons, and obtain the same behavior of not superimposing but very closely tracking.  As explained in Sec. I, for $q=2$ (Ising), only one type of random-field model exists.}
\end{figure}

In the bond-moving step of the transformation, $b^{d-1} = 4$ transfer matrices, where $b=2$ is the renormalization-group length-rescaling factor, are randomly chosen from the distribution and combined by multiplying the 4 elements at the same position in the 4 matrices, to form the bond-moved transfer matrix.  With no loss of generality, each element of the bond-moved transfer matrix is divided by the largest result of the fourfold multiplications, so the largest element of the bond-moved transfer matrix is 1 and the others are between 1 and 0.  This division is to avoid numerical blow-ups during the global renormalization-group flows.  This process is repeated until a new distribution of the bond-moved transfer matrices with $2q^2N$ elements is generated.

In the decimation step of the transformation, 2 transfer matrices are randomly chosen and matrix-multiplied.  The largest element is set to 1 by division as above.  This is repeated until $qN$ matrices are generated.  This distribution is symmetrized with respect to $<ij>$ by augmenting it with the transpose matrices and with respect to $q$ by augmenting it with the matrices where simultaneously each row and coulumn are cyclically augmented, again to cure any small asymmetry artificially introduced by the random numbers, reaching a distribution of renormalized transfer matrices with $2q^2N$ elements.  We have numerically found that $N=250$ is sufficient to obtain smooth results, giving up to $2q^2N =$ 24,500 elements (for $q=7$) in our distribution of transfer matrices.  This constitutes a single renormalization-group transformation.  The process is then repeated, starting with the bond-moving step.

\section{Results: 18 Phase Diagrams for the Random-Field Potts Models for $q$ = 2,3,4,5,6,7 States}
The calculated phase diagrams of the (3) random-axis randomly favored or disfavored random-field Potts models in $d=3$, for the number of states $q = 2,3,4,5,6,7$, are given Fig. 2, in terms of temperature $1/J$ and random-field strength $H/J$.  The $q=2$, namely Ising case, is as in Ref.\cite{Falicov}.  The phase diagrams of the random Potts models change with the number of states $q$.  Interestingly, the phase diagrams cross each other in proximity:  The ordered phases of the low-$q$ models reach higher temperatures, while those of the high-$q$ models reach higher random fields.  This has a physical explanation:  On the temperature axis, higher $q$ introduces more entropy and therefore lower free energy into the disordered phase where a spin visits all its states.  At low temperature in the high-random-field direction, spins differently pinned onto a higher number of $q-1$ nonordered states cannot create a correlated nonordered island among each other and are therefore less effective in disrupting order.

As discussed in Sec. I, the differently defined random-field Potts models: (1) Random-axis favored, (2) random-axis unfavored, (3) random-axis randomly favored or disfavored models, are differently defined models and are not expected to have superimposed phase diagrams for a given $q$. This is seen in Fig. 3, for $q = 3,5,7$, where Models (1),(2),(3) are not superimposed, but very closely track each other.  We have also calculated the phase diagrams for the three different random-field models for $q = 4,6$, not shown in Fig. 3 for congestion reasons, and see the same behavior of not superimposing but very closely tracking.  As explained in Sec. I, for $q=2$ (Ising), only one type of random-field model exists.
\section{Conclusion}
The random-field phase diagrams for $q=2,3,4,5,6,7$ state Potts models have been calculated in spatial dimension $d=3$.  As a function of increasing $q$, the ordered phase shows systematic excess or recess in the random-field or temperature direction, respectively.  This calculational results are explained by the differently pinned spins not being able to correlate into nonordering islands and by uncorrelated disordered spins visiting a larger number of states, respectively.  Similar behaviors can be expected in quenched-random symmetry-broken systems, as the simplicity of the Ising model is exceeded.

\begin{acknowledgments}
Support by the Academy of Sciences of Turkey (T\"UBA) is gratefully
acknowledged.
\end{acknowledgments}

\end{document}